\documentstyle[12pt]{article}
\begin{document} 
\vspace*{-1in} 
\renewcommand{\thefootnote}{\fnsymbol{footnote}} 
\begin{flushright} 
TIFR/TH/98-51\\
December 1998\\ 
hep-ph/9812486 
\end{flushright} 
\vskip 65pt 
\begin{center} 
{\Large \bf Large extra dimensions and deep-inelastic scattering at HERA}\\
\vspace{8mm} 
{\bf 
Prakash Mathews${}^{1}$\footnote{prakash@theory.tifr.res.in}, 
Sreerup Raychaudhuri${}^2$\footnote{sreerup@iris.hecr.tifr.res.in},   
K.~Sridhar${}^1$\footnote{sridhar@theory.tifr.res.in}
}\\ 
\vspace{10pt} 
{\sf 1) Department of Theoretical Physics, Tata Institute of 
Fundamental Research,\\  
Homi Bhabha Road, Bombay 400 005, India. 

2) Department of High Energy Physics, 
Tata Institute of Fundamental Research, \\  
Homi Bhabha Road, Bombay 400 005, India. } 
 
\vspace{80pt} 
{\bf ABSTRACT} 
\end{center} 
\vskip12pt 
In scenarios motivated by string theories, it is possible to have extra
Kaluza-Klein dimensions compactified to rather large magnitudes, 
leading to large effects of gravity at scales down to a TeV.
The effect of the spin-2 
Kaluza-Klein modes on the deep-inelastic cross-section at HERA
is investigated. We find that the data can be used to obtain bounds
on the effective low energy scale, $M_S$.
\setcounter{footnote}{0} 
\renewcommand{\thefootnote}{\arabic{footnote}} 
 
\vfill 
\clearpage 
\setcounter{page}{1} 
\pagestyle{plain}
\noindent The Standard Model (SM) has proved enormously successful in providing a 
description of particle physics upto energy scales probed by current
experiments, which is in the region of several hundred GeV. In the SM,
however, one assumes that effects of gravity can be neglected, because
the scale where the effects of gravity become large i.e. the Planck
scale ($M_P = 1.2 \times 10^{19}$~GeV) is vastly different from the TeV scale.
The separation between the TeV scale and the Planck scale is what 
manifests itself as the hierarchy problem, whose solution has become
one of the foci of the search for the correct physics beyond the SM.
This problem is exacerbated in traditional unification scenarios:
the scale of grand-unification is of the order of $10^{16}$~GeV and
again implies a huge desert. Further, in spite of the unification scale
being so close to the Planck scale, traditional unification models make
no reference whatsoever to gravity. 

Recent advances in string theory provide indications for a major paradigm
shift~-- in particular, unification of gravity with other interactions
now seems possible in the strongly-coupled limit of string theory called
$M$-theory \cite{hw, polchinski}. But other interesting effects may
also manifest at lower energies, for example, it was pointed out that 
the fundamental scale of string theory can be as low as a 
TeV \cite{lykken}. Of tremendous interest to phenomenology
is the possibility that the effects of gravity could become large at
very low scales ($\sim$~TeV), because of the effects of large extra
Kaluza-Klein dimensions where gravity can propagate \cite{dimo}. 
The starting point for such a scenario is a higher-dimensional theory
of open and closed strings \cite{dimo2, shiu}. The extra dimensions of 
this theory are
then compactified to obtain the effective low-energy theory in
3+1 dimensions, and it is assumed that $n$ of these extra dimensions
are compactified to a common scale $R$ which is relatively large,
while the remaining dimensions are compactified to much smaller
length scales which are of the order of the inverse Planck scale. 
In such a scenario, the SM particles correspond to open strings, which
end on a 3-brane. It implies that SM particles are localised on this 
3-brane, and are, therefore, confined to the $3+1$-dimensional spacetime. 
On the other hand, the gravitons (corresponding to closed strings) propagate 
in the $4+n$-dimensional bulk. The relation between 
the scales in $4+n$ dimensions and in $4$ dimensions is given by \cite{dimo}
\begin{equation} 
M^2_{\rm P}=M_{S}^{n+2} R^n ~,
\label{e1} 
\end{equation} 
where $M_S$ is the low-energy effective string scale. This equation has
the interesting consequence that we can choose $M_S$ to be of the order
of a TeV and thus get around the hierarchy problem. For such a value of
$M_S$, it follows that $R=10^{32/n -19}$~m, and so we find that $M_S$
can be arranged to be a TeV for any value $n > 1$. Effects of non-Newtonian
gravity can become apparent at these surprisingly low values of energy.
For example, for $n=2$ the compactified dimensions
are of the order of 1 mm, just below the experimentally tested region
for the validity of Newton's law of gravitation and within the possible
reach of ongoing experiments \cite{gravexp}.
In fact, it has been shown \cite{dimo4} that is 
possible to construct a phenomenologically viable scenario with large
extra dimensions, which can survive the existing astrophysical and 
cosmological constraints.

While the lowering of the string scale leads to the nullification of the 
hierarchy problem, the residual problem is that of stabilising the extra
large dimensions. This problem has been recently addressed in some papers
\cite{dimo3}. 
Moreover, the effect of the Kaluza-Klein states on the
running of the gauge couplings i.e. the effect of these states on the beta
functions of the theory have been studied \cite{dienes, kakushadze} and 
it has been shown that the unification scale can be also lowered down to scales 
close to the electroweak scale \footnote{Efforts to lower the compactification
scale have been made earlier in Ref.~\cite{anto}. The effects of Kaluza-Klein
states on the running of couplings was first investigated in 
Ref.~\cite{taylor}.}. For recent investigations on different aspects of the
TeV scale quantum gravity scenario and related ideas, see Ref.~\cite{related}.

Below the scale $M_S$ the following effective picture emerges \cite{sundrum, 
grw, hlz}: there are the Kaluza-Klein states, in addition to the usual
SM particles. The graviton corresponds to a tower of Kaluza-Klein states
which contain spin-2, spin-1 and spin-0 excitations. The spin-1
modes do not couple to the energy-momentum tensor and their
couplings to the SM particles in the low-energy effective
theory are not important. The scalar modes couple to the trace
of the energy-momentum tensor, so they do not couple to massless
particles. Other particles related to brane dynamics 
(for example, the $Y$ modes which are related to the
deformation of the brane) have effects which are subleading, compared to
those of the graviton. The only states, then, that contribute 
are the spin-2 Kaluza-Klein states. These
correspond to a massless graviton in the $4+n$ dimensional theory,
but manifest as an infinite tower of massive gravitons in the low-energy
effective theory. For graviton momenta smaller than the scale $M_S$, the
effective description reduces to one where the gravitons in the bulk 
propagate in the flat background and couple to the SM fields which live
on the brane via a (four-dimensional) induced metric $g_{\mu \nu}$. 
Starting from a linearized gravity Lagrangian
in $n$ dimensions, the four-dimensional interactions can be derived after
a Kaluza-Klein reduction has been performed. The interaction of the SM 
particles with the graviton, $G_{\mu\nu}$, can be derived from 
the following Lagrangian:
\begin{equation} 
{\cal L}=-{1 \over \bar M_P} G_{\mu \nu}^{(j)}T^{\mu\nu} ~,
\label{e2} 
\end{equation} 
where $j$ labels the Kaluza-Klein mode and $\bar M_P=M_P/\sqrt{8\pi}$,
and $T^{\mu\nu}$ is the energy-momentum tensor.
Using the above interaction Lagrangian the couplings of the graviton
modes to the SM particles can be calculated \cite{grw,hlz}, and
used to study the consequences at colliders of this TeV scale 
effective theory of gravity. In particular, direct searches for graviton 
production at $e^+ e^-$, $p \bar p$ and $pp$ colliders, leading to 
spectacular single photon + missing energy or monojet + missing energy 
signatures, have been suggested \cite{grw, mpp, hlz}. The virtual effects 
of graviton exchange in $e^+ e^- \rightarrow f \bar f$ and in high-mass 
dilepton production \cite{hewett}, and in $t \bar t$ production \cite{us} 
at the Tevatron and the LHC have been studied. The bounds on $M_S$ obtained 
from direct searches depend on the number of extra dimensions. Non-observation
of the Kaluza-Klein modes yield bounds which are around 500 GeV to 1.2 TeV 
at LEP2 and around 600 GeV to 750 GeV at Tevatron (for $n$ between 
2 and 6) \cite{mpp}. Indirect bounds from virtual graviton exchange
in dilepton production at Tevatron yields a bound of around 950 GeV 
\cite{hewett}. Virtual effects in $t \bar t$ production at Tevatron 
yields a bound of about 650 GeV \cite{us}.

In view of the
fact that the effective Lagrangian given in Eq.~\ref{e2} is suppressed
by $1/\bar M_P$, it may seem that the effects at colliders will be hopelessly
suppressed. However, in the case of real graviton production, the phase
space for the Kaluza-Klein modes cancels the dependence on $\bar M_P$ 
and, instead, provides a suppression of the order of $M_S$. For the
case of virtual production, we have to sum over the whole tower of 
Kaluza-Klein states and this sum when properly evaluated \cite{hlz, grw}
provides the correct order of suppression ($\sim M_S$). The summation
of time-like propagators and space-like propagators yield exactly the
same form for the leading terms in the expansion of the sum \cite{hlz}
and this shows that the low-energy effective theories for the $s$ and 
$t$-channels are equivalent.

In the present work, we study the effect of the virtual graviton
exchange on the $e^+ p$ deep-inelastic scattering cross-section 
at HERA. The presence of the new couplings from the low-energy effective
theory of gravity, lead to new $t$-channel diagrams in the 
$e^+ q(\bar q)$ or $e^+ g$ initial state. We use the couplings
as given in Refs.~\cite{grw, hlz}, and summing over all the graviton modes,
we find the following expressions for the
cross-sections involving the virtual graviton exchange (in the following
we use the notation $d\hat \sigma^{a(i)}/d\hat t$, for the process
$e^+ i \rightarrow e^+ i$, where $i=q,g$ is the parton in the initial
state, and the process is mediated by the exchange of $a$; and
the notation $d\hat \sigma^{ab(i)}/d\hat t$, for the interference of 
processes $e^+ i \rightarrow e^+ i$ mediated by the exchanges of the
virtual particles $a$ and $b$, respectively):

\begin{eqnarray}
{{\rm d} \hat \sigma (e^+q \rightarrow e^+q) \over {\rm d} \hat t}&=&
{{\rm d} \hat \sigma^{\rm SM} \over {\rm d} \hat t}+
{{\rm d} \hat \sigma^{G(q)} \over {\rm d} \hat t}+
{{\rm d} \hat \sigma^{\gamma G(q)} \over {\rm d} \hat t}+
{{\rm d} \hat \sigma^{ZG(q)} \over {\rm d} \hat t}, \\
{{\rm d} \hat \sigma^{G(q)} \over {\rm d} \hat t}&=&
{\pi \lambda^2 \over 32 M_S^8}  {1 \over \hat s^2}
\left[32 \hat u^4 + 64 \hat u^3 \hat t + 42 \hat u^2 \hat t^2 + 
10\hat u \hat t^3 + \hat t^4\right] ,\\
{{\rm d} \hat \sigma^{\gamma G(q)} \over {\rm d} \hat t}&=&
{\pi \alpha e_q \lambda \over 2 M_S^4}  ~\frac{1} 
{\hat s^2 \hat t} ~(2 \hat u + \hat t)^3 ,\\
{{\rm d} \hat \sigma^{ZG(q)} \over {\rm d} \hat t}&=&
{\pi \alpha \lambda \over 2 \sin^2 2\theta_w M_S^4} 
{\left[C_V^e C_V^q (2 \hat u+\hat t)^3 + C_A^e C_A^q (6 \hat u^2+6 \hat u 
\hat t+\hat t^2)\right]
\over \hat s^2 (\hat t-m_Z^2)}
, \\
{{\rm d} \hat \sigma(e^+g \rightarrow e^+ g) \over {\rm d} \hat t}&=&
{{\rm d} \hat \sigma^{G(g)} \over {\rm d} \hat t}=
{\pi \lambda^2 \over 2 M_S^8} ~{\hat u \over \hat s^2}~
\left[2 \hat u^3 +4 \hat u^2 t + 3 \hat u \hat t^2 + \hat t^3 \right] ,
\end{eqnarray}
$\lambda$ is the coupling at the effective scale $M_S$ and is expected 
to be of ${\cal O}(1)$, but its sign is not known $a\ priori$ and
$C_V=T^3 -2{\rm sin}^2\theta_W Q_f$ and $C_A=T_3$ are 
the usual vector and axial-vector couplings of the
fermions to the $Z$. 
In our work we will explore the sensitivity of our results to the 
choice of the sign of $\lambda$. The $e^+ g$ subprocess is, of course,
absent in the SM, and is completely a result of introducing the new
interactions. The new interactions also contribute in the $e^+ q$
channel, where there is an interference between the SM amplitude
and the amplitude due to the new physics. 

%========================================================================%
\begin{figure}[ht]
\begin{center}
\vspace*{4.0in}
      \relax\noindent\hskip -5.2in\relax{\includegraphics{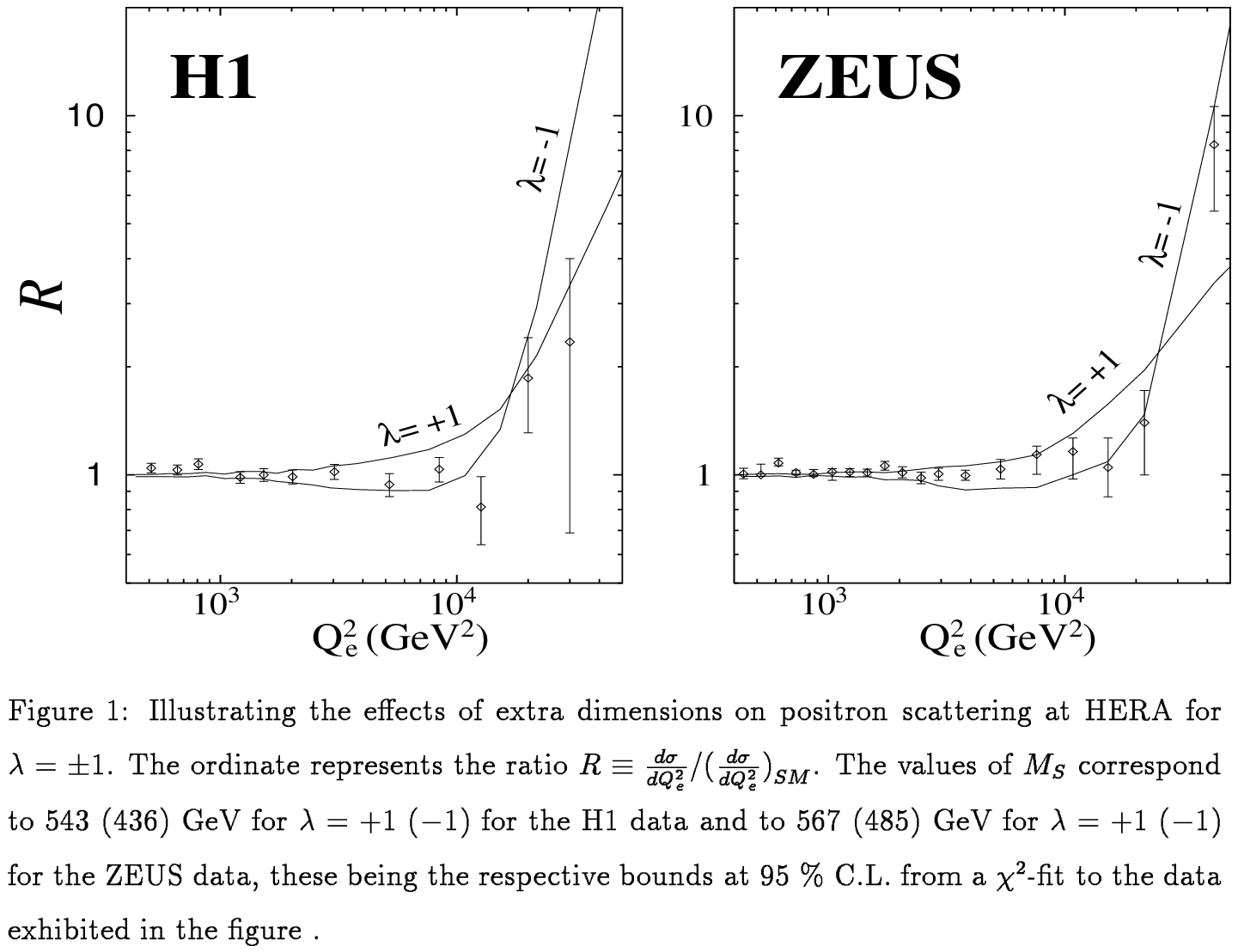}}
\end{center}
%\vspace{+1.0in}
%\caption{\footnotesize\it Illustrating the variation of the $t \bar t$
%                          cross-section with variation in the scale
%                          $M_S$ at (a) the Tevatron and (b) the LHC.
%                          For the Tevatron, dashed lines show the 
%                          experimental data from Run I, when the CDF
%                          and D0 results are combined. For the LHC,
%                          the dashed lines correspond to errors on
%                          the SM cross-section of 0.3, 0.5 and 1 pb 
%                          respectively. } 
\end{figure}
%========================================================================%

The H1 \cite{h1} and ZEUS \cite{zeus} collaborations at HERA have
presented their results from their combined 1994-97 runs in terms of
the quantity
\begin{equation}
R={d\sigma/dQ^2_e \vert_{\rm exp} \over d\sigma/dQ^2_e \vert_{\rm SM}},
\end{equation}
where $Q^2_e$ is the momentum-transfer squared constructed from the
$e^+$ track. In our parton-level simulation, $Q^2_e \sim -\hat t$. 
We use this observable to compare against 
\begin{equation}
R={d\sigma/dQ^2_e \vert_{\rm SM+NSM} \over d\sigma/dQ^2_e \vert_{\rm SM}},
\end{equation}
where NSM denotes non-Standard Model, and use the data to put bounds
on the value of $M_S$. The cross-section $d\sigma/dQ^2_e$ is given as
\begin{equation}
{d\sigma (e^+p \rightarrow e^+ jet) \over dQ^2_e}= \sum \int dx
f_{i/p}(x) {d\hat \sigma \over d\hat t} ,
\label{e5}
\end{equation}
$f_{i/p}$ denotes the probability of finding a parton $i$ in the proton.
The sum in Eq.~\ref{e5} runs over the contributing subprocesses.

The results of our numerical evaluation of the cross-section are shown
in Figure 1. We have plotted $R$ as a function of $Q^2_e$, as obtained 
from our calculation and compared it with each experiment separately.
For our computations, we use the cuts as used in the two experiments
and we have used CTEQ4 parton densities \cite{cteq} taken from PDFLIB
\cite{pdflib}.
The curves represent the 95\% C.L. bounds from a $\chi^2$ fit to each
set of data. These fits yield the bound on $M_S$ to be 543 (436) GeV
for $\lambda= +1 (-1)$ for the H1 data and 567 (485) GeV
for $\lambda= +1 (-1)$ for the ZEUS data. For higher values of MS,
the non-Standard contribution decreases, so that the
curves move closer to $R=1$.

The data on $R$ from H1 and ZEUS collaborations, in fact, show a deviation
from the SM for $Q^2_e$ values beyond $10^4$~GeV${}^2$. The errors on the
ratio $R$ at these large values of $Q^2_e$ are large, so this discrepancy
with the SM prediction is not very significant statistically. The result of
including the non-Standard Model contribution is to improve the $\chi^2$
of our fit to the data. We find that for $\lambda=-1$, the theoretical
prediction, with the non-Standard Model contribution, fits the data
rather well, even accounting for the modest dip in $R$ at a value of
$Q^2_e$ just below $10^4$ GeV${}^2$! This behaviour follows from the fact that
the interference term in the quark-initiated sector dominates at relatively 
low $Q^2_e$, and this gives a negative contribution for $\lambda=-1$. As
one moves to larger $Q^2_e$, the gluon-initiated contribution starts to
dominate and gives the increase at large $Q^2_e$. Given that a discrepancy
with the SM seen in the experiments exists, though it is not 
statistically compelling, the bounds we derive from the data are
not as strong as compared to those derived from Tevatron data on dilepton
production \cite{hewett} and $t \bar t$ production \cite{us}. In the
event that the HERA experiments improve their data in the large $Q^2_e$
region and find good agreement with the SM, the bounds presented in
this paper are likely to improve considerably. For example, with the
20 fold increase in luminosity planned in HERA experiments in the
next few years, assuming that the data are centred around the Standard
Model prediction we estimate that the bounds on $M_S$ would go up to 
around 600 (925) GeV for $\lambda= +1 (-1)$. 

We have studied the effect of large extra dimensions and a TeV scale
gravity on the deep inelastic scattering cross-section at HERA. 
The fits of the theoretical curves to the data yield the value of 
the effective string scale, $M_S$, to be $>$~543 (436) GeV
for $\lambda= +1 (-1)$ for the H1 data and $>$~567 (485) GeV
for $\lambda= +1 (-1)$ for the ZEUS data. The bounds are likely to
increase with any improvement in the data, especially at large $Q^2_e$.

\end{document}